# A method for determining the transition energies of $^{83m}$Kr at the KATRIN experiment

C. Rodenbeck

Institut für Kernphysik
Westfälische Wilhelms-Universität Münster
Wilhelm-Klemm-Str. 9, 48149 Münster, Germany

May 18, 2022

The neutrino mass experiment KATRIN uses conversion electrons from the 32.2-keV transition of the nuclear isomer $^{83m}$Kr for calibration. Comparing the measured energies to the appropriate literature values allows for an independent evaluation of the energy scale, but the uncertainties in some of the literature values obtained by gamma spectroscopy are a limiting factor. Building upon the already excellent linearity of KATRIN's energy scale, this paper proposes a novel method for determining the $^{83m}$Kr transition energies via high-precision electron spectroscopy. Notably, the method makes use of conversion electrons from the 41.6-keV direct transition of $^{83m}$Kr to its ground state in addition to conversion electrons from the much more frequent cascade of a 32.2-keV and a 9.4-keV transition. By implementing this method, KATRIN may be able to deliver order-of-magnitude improvements in precision over current $^{83m}$Kr transition energy literature values.

# 1 Introduction

The Karlsruhe Tritium Neutrino (KATRIN) experiment measures the effective mass of the electron anti-neutrino [1, 2]. A sensitivity of 0.2 eV/c$^2$ (at 90 % confidence level) is targeted after collecting three live years of data.

The KATRIN setup consists of a high-luminosity tritium source and a high-precision integrating spectrometer that combines magnetic collimation and an electrostatic filter [3–5]. Beta-decay electrons from the Windowless Gaseous Tritium Source (WGTS) are guided through the transport and pumping section [6] to the spectrometer and detector section [7]. The detector is reached only by those electrons whose kinetic energies are large enough to overcome the retarding potential inside the main spectrometer. An integrated electron spectrum is accumulated by step-wise adjustments of the retarding potential.

For the neutrino mass determination, the tritium spectrum is measured around its endpoint at 18.6 keV, where the imprint of the neutrino mass is largest. To avoid a bias on the neutrino mass, KATRIN's energy scale needs to be linear and stable during measurements.

In the tritium spectrum fit of the neutrino mass analysis, the squared neutrino mass value and the effective endpoint are strongly correlated parameters (0.97 [8]). Any unknown systematic leading to a shifted effective endpoint also biases the measured neutrino mass. For example, a Gaussian broadening of the energy scale of 60 meV would shift the measured squared neutrino mass value by $7 \times 10^{-3}$ meV$^2$ [1].

The steadiness of the fitted effective endpoint of the tritium spectrum is a good proxy for a stable energy scale (cf. [9]), but the absolute energy scale also needs to be known: As outlined in the work [10] and since implemented at KATRIN [8], an important check for systematics is to translate the effective tritium endpoint value into the Q-value for comparison with independent measurements of the tritium–helium-3 mass difference [11, 12].

# 2 Energy scale calibration at KATRIN using $^{83m}$Kr

For the absolute calibration of KATRIN's energy scale, monoenergetic conversion electrons from $^{83m}$Kr are used as a reference.

$^{83m}$Kr usually decays in a cascade of two transitions (32.2 keV, 9.4 keV) to $^{83}$Kr. These transitions can take place in the form of gamma emissions, or, more frequently, in the form of internal conversion electron emissions. The energy $E$ released in each alternate process is

$$E = E_\gamma + E_{\gamma\,\text{rec}} = E_{\text{ce}} + E_{\text{bind}} + E_{\text{rec}}, \qquad (1)$$

where $E_\gamma$ denotes the energy of the emitted gamma ray, $E_{\text{ce}}$ the kinetic energy of the conversion electron, and $E_{\text{bind}}$ its binding energy. $E_{\gamma\,\text{rec}}$ and $E_{\text{rec}}$ are the recoil energies of the atom after emission of a gamma ray or a conversion electron, respectively. $E$ is hereafter called simply the *transition energy*.

As follows from eq. (1), the kinetic energy $E_{ce}$ of a $^{83m}$Kr conversion electron expressed in terms of the other energies is

$$E_{ce} = E_\gamma + E_{\gamma\,rec} - E_{bind} - E_{rec}$$
$$= E - E_{bind} - E_{rec}. \qquad (2)$$

Literature values are available for $E_\gamma$, $E_{\gamma\,rec}$, $E_{bind}$, and $E_{rec}$ (cf. [13]).

The $^{83m}$Kr conversion electron energy $E_{ce}$ can be measured at KATRIN. For this, a $^{83}$Rb source emanating $^{83m}$Kr (and $^{83}$Kr) [14] is attached to the gas circulation sytem of the WGTS, transforming it into a gaseous krypton source (GKrS). Additionally, a condensed krypton source (CKrS) [15] is available, located between the transport and pumping section and the spectrometer section. The work [16] describes in detail how the position (i.e., energy) and width of the conversion electron line are determined by fitting the integrated spectrum. The line position in the measured spectrum deviates from the true value of $E_{ce}$ by the shift $\Delta\Phi'$ of KATRIN's absolute energy scale at the time of the measurement.

## 2.1 Shift of the energy scale

There are many possible sources along the beamline that can impact the energy scale and contribute to the shift $\Delta\Phi'$. Most of them are known very well, and methods exist to characterize them.

On the spectrometer side, the main spectrometer's work function $\Phi_{MS}$ and its retarding potential influence the energy scale. The work function can be measured with an electron gun [17, 18]. The retarding potential is kept stable within 2 ppm ($2 \times 10^{-6}$) by the precision high-voltage setup [19]. The scale factor of the high-voltage dividers [20, 21] measuring the retarding potential $U_{ret}$ is known from the absolute calibration [22], with systematic uncertainties within 1 ppm. The retarding potential depression $U_{dep}$[1] is known from simulations, with an uncertainty $\delta U_{dep}$.

On the source side, the energy scale is defined by the starting potential $U_{start}$ of the electrons and their energy loss due to scattering inside the source. The work function of the WGTS, the work function of the rear wall[2] and its bias voltage, and the plasma potential govern the starting potential. The stability and spatial distribution of the starting potential inside the WGTS are investigated with $^{83m}$Kr measurements [23]. The energy loss spectrum needs to be known precisely for measuring the continuous tritium spectrum and it is characterized with electron gun measurements [24]. For the monoenergetic conversion electron lines, the scattering process only reduces the intensity of the measured signal (the unscattered electrons), but does not change the measured line position.

---

[1]The potential depression is the difference between the retarding voltage applied at the spectrometer vessel and the actual potential seen by the electrons inside the spectrometer.

[2]The rear wall [2] is a gold coated disc at the far end of the WGTS which absorbs most of the tritium decay electrons which were not transmitted.

Considering all contributions and including $U_{\text{ret}}$ and $U_{\text{dep}}$ in the conversion electron spectrum fit leads to a measured line position of

$$E_{\text{M}} = q \cdot (U_{\text{ret}} - U_{\text{dep}}) = E_{\text{ce}} + q \cdot \delta U_{\text{dep}} + q \cdot U_{\text{start}} + \Phi_{\text{MS}} \qquad (3)$$
$$= E_{\text{ce}} + \Delta \Phi',$$

where $q$ is the charge of the electron and $\Delta\Phi'$ includes all unknowns $\delta U_{\text{dep}}$, $\Phi_{\text{MS}}$ and $U_{\text{start}}$. It should be noted that the particular sources of the individual shifts are not important for the conversion electron measurements. All of them combine into one effective shift of the measured line position relative to the actual value.

Conversion electron spectra of the 32-keV transition are regularly measured at KATRIN and reveal the temporal variation of $\Delta\Phi'$. Additionally, by comparing the [83m]Kr line positions to the literature values, $\Delta\Phi'$ can be determined (cf. eq. (3)) and with it KATRIN's absolute energy scale.

## 3 Method for determining the transition energies

When comparing the measured conversion electron lines to the literature, the dominant contributor of uncertainty is $E_\gamma$, which is known to 0.5 eV. KATRIN, meanwhile, has been shown to be able to perform electron spectroscopy with a precision of 0.025 eV [16], and future performance improvements are plausible. Despite its high resolution and the excellent linearity of its energy scale [16, 22], KATRIN's accuracy is limited by the uncertainty in $\Delta\Phi'$. The method presented here first removes this limitation so that the transition energies can then be determined with high precision.

### 3.1 Leveraging the cross-over transition

It is possible to determine $\Delta\Phi'$ *independently* of any transition energy literature value by measuring conversion electron lines from a set of three interconnected transitions: the 32-keV and 9-keV cascade transitions, as well as the corresponding 42-keV cross-over transition[3] which can occur in their stead.

The transition energies $E(g)$ of each of the three transitions $g \in \{\tau_{32}, \tau_9, \tau_{42}\}$ relate to each other as

$$E(\tau_{42}) = E(\tau_{32}) + E(\tau_9). \qquad (4)$$

Generalizing and simplifying eqs. (2) and (3) leads to a measured line position

$$E_{\text{M}}(g, s) = E(g) - E_{\text{bind}}(s) - E_{\text{rec}}(g) + \Delta\Phi', \qquad (5)$$

---
[3]The 42-keV transition is highly suppressed and no precision spectroscopy of its conversion electrons has been performed yet. The electrons are however visible in the detector spectrum during routine [83m]Kr measurements at KATRIN (discussed in e.g. [25]).

with $s \in \{K, L_3, N_2, ...\}$ denoting the conversion electron's subshell.

Arbitrary subshells $s_{32}, s_9, s_{42}$ can be chosen for the measurement of each respective transition. Inserting eq. (5) into eq. (4), the shift is then determined as

$$\begin{aligned}\Delta\Phi' = &- E_M(\tau_{42}, s_{42}) - E_{\text{bind}}(s_{42}) - E_{\text{rec}}(\tau_{42}) \\ &+ E_M(\tau_{32}, s_{32}) + E_{\text{bind}}(s_{32}) + E_{\text{rec}}(\tau_{32}) \\ &+ E_M(\tau_9, s_9)\ \ \ + E_{\text{bind}}(s_9)\ \ + E_{\text{rec}}(\tau_9)\ .\end{aligned} \quad (6)$$

It is important to note that when determined in this way, $\Delta\Phi'$ does not depend on any transition energy $E$, and by extension does not depend on any gamma emission energy $E_\gamma$.

For eq. (6) to be valid, $\Delta\Phi'$ of course needs to be the same for all three measurements. A constant $\Delta\Phi'$ can be achieved by measuring under the same conditions along KATRIN's beamline. This effectively means measuring in direct succession to ensure unchanged source conditions. The retarding potential at the main spectrometer, meanwhile, necessarily changes as part of the measuring process. The absolute calibration method [22] however characterizes the voltage dependency of the high-voltage divider measuring the retarding potential with a precision of 1 ppm. Furthermore, the systematic error $\delta U_{\text{dep}}$ of the main spectrometer field simulations to determine $U_{\text{dep}}$ is independent of the voltage value. To summarize, $\Delta\Phi'$ is constant within an acceptable and known margin of error despite the varying retarding potential.

## 3.2 Calculating the transition energies

By solving each of the measurement series' instances of eq. (5) for $E(g)$ and inserting $\Delta\Phi'$ as defined by eq. (6), the transition energies can be determined:

$$\begin{aligned}E(g) = &\ E_M(g, s) + E_{\text{bind}}(s) + E_{\text{rec}}(g) - \Delta\Phi' \\ = &\ E_M(g, s) + E_{\text{bind}}(s) + E_{\text{rec}}(g) \\ &+ E_M(\tau_{42}, s_{42}) + E_{\text{bind}}(s_{42}) + E_{\text{rec}}(\tau_{42}) \\ &- E_M(\tau_{32}, s_{32}) - E_{\text{bind}}(s_{32}) - E_{\text{rec}}(\tau_{32}) \\ &- E_M(\tau_9, s_9)\ \ \ - E_{\text{bind}}(s_9)\ \ - E_{\text{rec}}(\tau_9)\ .\end{aligned} \quad (7)$$

The uncertainties in the recoil energies [13] are negligible here, leaving only the uncertainties in the electron's binding energies and those in the measured line positions. As can be seen in eq. (7), using the same subshell for all measurements (i.e., $s_{32} = s_9 = s_{42}$) has the advantage of the binding energies canceling each other out, resulting in the simpler form

$$\begin{aligned}E(g) = &\ E_M(g, s) + E_{\text{rec}}(g) \\ &+ E_M(\tau_{42}, s_{42}) + E_{\text{rec}}(\tau_{42}) \\ &- E_M(\tau_{32}, s_{32}) - E_{\text{rec}}(\tau_{32}) \\ &- E_M(\tau_9, s_9)\ \ \ - E_{\text{rec}}(\tau_9)\ .\end{aligned} \quad (8)$$

Table 1: Energies (in eV) involved in $^{83\mathrm{m}}$Kr gamma decay and internal conversion for different transitions $g$ and a selection of subshells $s$. Adapted from [13].

| | $E_\gamma$ | $E_{\gamma\,\mathrm{rec}}$ | | $E_\mathrm{bind}$ | $E_\mathrm{rec}$ | | |
| --- | --- | --- | --- | --- | --- | --- | --- |
| | | | | | $g = \tau_{32}$ | $g = \tau_9$ | $g = \tau_{42}$ |
| $g = \tau_{32}$ | 32 151.6(5) | 0.006 7 | $s = \mathrm{K}$ | 14 327.26(4) | 0.120 | – | 0.185 |
| $g = \tau_9$ | 9 405.7(6) | 0.000 57 | $s = \mathrm{L}_3$ | 1 679.21(5) | 0.207 | 0.051 | 0.274 |
| | | | $s = \mathrm{N}_2$ | 14.67(1) | 0.219 | 0.063 | 0.286 |
| | | | $s = \mathrm{N}_3$ | 14.00(1) | 0.219 | 0.063 | 0.286 |

For an exemplary uncertainty estimation, let's assume a measurement series using exclusively electrons from the $\mathrm{L}_3$ subshell. The $\mathrm{L}_3$-32 line measurement is routine at KATRIN. The latest published result of the line position with the GKrS has a statistical uncertainty of 3 meV and a systematic uncertainty of 25 meV [16]. The $\mathrm{L}_3$-9 measurement is not a routine measurement, and the $\mathrm{L}_3$-42 line has yet to be measured at all. The challenges for both measurements are discussed in section 5. Assuming that the challenges are overcome, one could expect the same uncertainty as for the $\mathrm{L}_3$-32 line measurement (25 meV). Under this assumption, the energy of the 32-keV transition can be determined with an uncertainty of 35 meV, a considerable improvement over the 500 meV uncertainty in the current literature values (cf. table 1).

For the 9-keV transition, the expected uncertainty would be the same. Since the 42-keV transition is the sum of the other two transitions, the uncertainties of all three line positions contribute (cf. eq. (8)), leading to an uncertainty of 50 meV.

A proper uncertainty evaluation is of course only possible after the measurements have been performed. The values should be understood as an estimate of what may be achievable with this method.

## 4 Transfering the gains to routine calibrations

The higher-precision $^{83\mathrm{m}}$Kr transition energies can now serve as a better calibration reference for KATRIN's energy scale when using only a single $^{83\mathrm{m}}$Kr conversion electron line (cf. section 2.1).

One example is a simple one-hour measurement of the $\mathrm{L}_3$-32 line position. To match the conditions of the neutrino mass measurements as closely as possible, tritium and $^{83\mathrm{m}}$Kr should be circulated together inside the WGTS as was done in [23]. With the $\mathrm{L}_3$-32 line position measurement, the shift can be determined as (cf. eq. (5))

$$\Delta\Phi' = E_\mathrm{M}(\tau_{32}, \mathrm{L}_3) - E(\tau_{32}) + E_\mathrm{bind}(\mathrm{L}_3) + E_\mathrm{rec}(\tau_{32}). \tag{9}$$

Following the assumption that all challenges described in section 5 are met to successfully determine $E(\tau_{32})$ with a precision of 35 meV (cf. section 3.2), the total uncertainty in $\Delta\Phi'$ would be 66 meV. Since the $L_3$-32 line position measurements are performed regularly, any improvements can also be applied retroactively to $\Delta\Phi'$. With high source activity, the measurement of the $N_2N_3$-32 doublet is also feasible, leading to an even smaller error in $\Delta\Phi'$ due to the smaller uncertainty in the binding energy (cf. table 1).

By measuring $\Delta\Phi'$ during a neutrino mass measurement campaign, $\Delta\Phi'$ can be applied to the effective endpoint of tritium spectrum fits and translated into the measured Q-value.

To put the uncertainty of $\Delta\Phi'$ in perspective, the Q-value is known to 0.07 eV from external measurements [12] while the uncertainty is 0.6 eV in the most recent KATRIN publication [23].

## 5 Challenges

An ideal measurement would use both the CKrS and the GKrS at the KATRIN beamline. Comparing the resulting transition energies can give access to hidden systematics. However, the 9-keV transition can be challenging to measure with the GKrS.

The first complication is that the conversion electron lines of the 9-keV transition are below the tritium endpoint. This complication is easy to solve by using helium as carrier gas, or circulating only krypton inside the source as was done in [16]. Then the background due to tritium is reduced to only residual tritium on the rear wall, which can be minimized by cleaning.

The second complication is that the 32-keV transition leaves the atom in a multiply-ionized state, and the 9-keV transition follows the 32-keV transition within the 155 ns half-life [26]. Due to the low $^{83m}$Kr density inside the WGTS, the neutralization times are longer than the half-life. The multiply-ionized states shift the binding energies. Instead of one conversion electron line, multiple shifted lines are visible, with each corresponding to a charged state [27]. For a successful measurement, the charged states need to be identified. The work [28] describes the first measurement of this kind at KATRIN. Inside the CKrS, the neutralization times are shorter than the half-life, thus no multiply-ionized states exist at the time of the 9-keV transition. Therefore the observed conversion electron energy is not shifted.

For the CKrS, the binding energies of the different shells have an unknown shift due to condensed matter effects. Here it is vital that the three lines all come from the same subshell, as was done in eq. (8), to cancel out the unknown binding energy.

Another challenge will be the measurement of the highly suppressed 42-keV transition. The expected intensity of the $L_3$-42 line is roughly 13 ppm of the $L_3$-32 line, and roughly 0.16 % of the $N_3$-32 line [29]. However, the lack in intensity can be overcome by a high-luminosity source and an increase in measurement time. For example, successful high-intensity measurements of the $N_2N_3$-32 line doublet were already performed with a 10 GBq GKrS. The energetically higher lines from the 42-keV transition were clearly visible as background.

To measure the integrated spectrum of the $L_3$-42 line, a voltage of around $-39.8$ kV needs to be applied at the KATRIN spectrometer. During normal operation only voltages down to $-35$ kV are permitted. Several, but feasible, hardware changes are needed to operate the spectrometer at voltages down to $-40$ kV. The key to the precision high-voltage setup is the divider, and fortunately, one of the two available dividers is designed to measure voltages up to $\pm 65$ kV [21].

When using the GKrS,[4] measuring the K-42 line at 27.2 keV instead is also a possibility. Since the spectrometer acts as a high pass filter, the signal would be lost in the background of the energetically higher lines (e.g. the $L_3$-32 at 30.5 keV has a roughly $4 \times 10^4$ higher intensity). Changing the magnetic fields to reduce the transmittance of electrons with high-surplus energy, or improving the energy resolution at the detector are possible solutions.

## 6 Conclusion

It has been shown that by measuring three distinct transitions of $^{83m}$Kr, the shift $\Delta\Phi'$ of KATRIN's energy scale can be determined independently of the absolute transition energies. This allows the $^{83m}$Kr transition energies with a precision improved over the current literature values to be determined. These higher-precision transition energy values translate directly into more precise single-line $^{83m}$Kr calibration measurements at KATRIN, even retroactively.

## Acknowlegdements

I would like to thank F. Fränkle, V. Hannen, A. Lokhov, A. Marsteller, A. Poon, R. G. H. Robertson, M. Schlösser, T. Thümmler, D. Vénos, and C. Weinheimer for the helpful discussions. This work was supported by the BMBF under reference 05A20PMA and by the DFG through the Research Training Network 2149.

---

[4]It is not possible to use the K-42 line with the CKrS since all measurements need to be from the same subshell, as explained earlier. The 9-keV transition is not possible for electrons from the K shell due to their binding energies of about 14 keV.